\documentclass{article}

\usepackage[preprint,nonatbib]{neurips_2020}

\usepackage[utf8]{inputenc} %
\usepackage[T1]{fontenc}    %
\usepackage[russian, english]{babel}
\usepackage{hyperref}       %
\usepackage{url}            %
\usepackage{booktabs}       %
\usepackage{amsfonts}       %
\usepackage{nicefrac}       %
\usepackage{microtype}      %
\usepackage{graphicx}
\usepackage{todonotes}
\usepackage{enumitem}
\usepackage{csquotes}
\usepackage{xcolor}
\usepackage{xspace}
\newcommand\mypara[1]{\vspace{0.5mm}\noindent\textbf{#1}}

\makeatletter
\DeclareRobustCommand\onedot{\futurelet\@let@token\@onedot}
\def\@onedot{\ifx\@let@token.\else.\null\fi\xspace}

\def\etal{\emph{et al}\onedot}

\definecolor{blue(pigment)}{rgb}{0.2, 0.2, 0.6}
\definecolor{green(pigment)}{rgb}{0.1, 0.5, 0.1}

\newcommand{\dmytro} [1] {\textcolor{green(pigment)}{\textbf{Dmytro:} #1}}

\renewcommand{\dmytro} [1] {#1}

\title{ArXiving Before Submission Helps Everyone}

\author{%
  Dmytro Mishkin\thanks{} \\
  Visual Recognition Group,\\ Faculty of Electrical Engineering,\\ Czech Technical University in Prague\\
  \texttt{ducha.aiki@gmail.com} \\
  \And
  Amy Tabb \\
  writing in her personal capacity\\
  U.S.A \\
  \texttt{amy@amytabb.com} \\
    \And
  Jiri Matas \\
  Visual Recognition Group, \\Faculty of Electrical Engineering, \\Czech Technical University in Prague\\
  \texttt{matas@fel.cvut.cz} \\

}

\begin{document}

\maketitle

\begin{abstract}
We claim, and present evidence, that allowing arXiv publication before a conference or journal submission benefits researchers, especially early career, as well as the whole scientific community. 
Specifically, arXiving helps professional identity building, protects against independent re-discovery, idea theft and gate-keeping; it facilitates open research result distribution and reduces inequality.
The advantages dwarf the drawbacks -- mainly the relative increase in acceptance rate of papers of well-known authors -- which studies show to be marginal. Analyzing the pros and cons of arXiving papers, we conclude that requiring preprints be anonymous is nearly as detrimental as not allowing them. We see no reasons why anyone but the authors should decide whether to arXiv or not.

\end{abstract}

\section{Introduction}
\label{sec:intro}
Releasing research papers on the arXiv open repository~\cite{arxiv} is becoming a standard in computer science and other fields ~\cite{biorxiv, medrxiv, psyarxiv}. We argue that arXiving provides great benefits to both individual researchers (Section~\ref{sec:researcher}) and a field as a whole (Section~\ref{sec:field}). 
Moreover, it helps researchers in all career stages; we fail to find circumstances in which researchers do not benefit from arXiving their work.  

There are opinions, however, that this open science practice should be prohibited or limited to arXiving anonymized papers only. The beliefs underpinning these views are the following: 1. waiting 3-6 months until a paper is accepted at a conference is a minor inconvenience, and  2. arXiv releases lead to an "almost" single blind conference review process, which may reduce diversity and some studies~\cite{Tomkins12708} claim that it favours people from well-known institutions. %

We address the first contra argument in Section~\ref{sec:researcher} and 
single blind review issues in Section~\ref{sec:fairness-single}.

\section{Individual researchers benefit from pre-submission arXiving}
\label{sec:researcher}

We start by discussing the benefits of arXiv to, especially, early career researchers (ECRs), from the opportunities that come from \textbf{early released non-anonymous preprints}. We will call such practice "arXiving" in the rest of the paper.

\mypara{Professional identity building instead of blank C.V.} 
When arXiving early, 
authors, particularly ECRs, can build professional identities around the work instead of waiting until acceptance. Activities include events, collaborations, and building informal networks of support outside of their institution~\cite{montgomery2018}. All scholarly work can be listed on a C.V. -- including preprints –- to show research output. While having the most recent publication online might not be important for a professor with 200+ publications, for the early career researcher it is the step from zero to one. The same is true for the time -- 6 months can make career life-or-death difference for an ECR. 

Increased early visibility through preprints allows others to become aware of new authors. 
For instance recruiters, who, in our experience, contact candidates; it is not candidates sending applications. Other examples include fellow researchers who are looking for speakers at local meetups, non-academic conferences, workshops, departmental seminars, and so on.

In the "wait until acceptance" scenario, an ECR could end up with a blank C.V. \emph{for a few years}.
This is even more significant now, when publications are required to \emph{enter the PhD program}.

\mypara{Protection against independent re-discovery or idea theft.} Even if a paper is well-written and has a useful contribution, there is still a high chance that it will be rejected. 
The review process, even at the best conferences, is known to be quite random~\cite{nips-experiment}. 
Low acceptance rates of 20-30\%~\cite{conf-stats} at CVPR and NeurIPS-like top conferences exacerbate the decision noise problem. %

\begin{figure}[tpb]
\centering
 \includegraphics[width=0.32\linewidth]{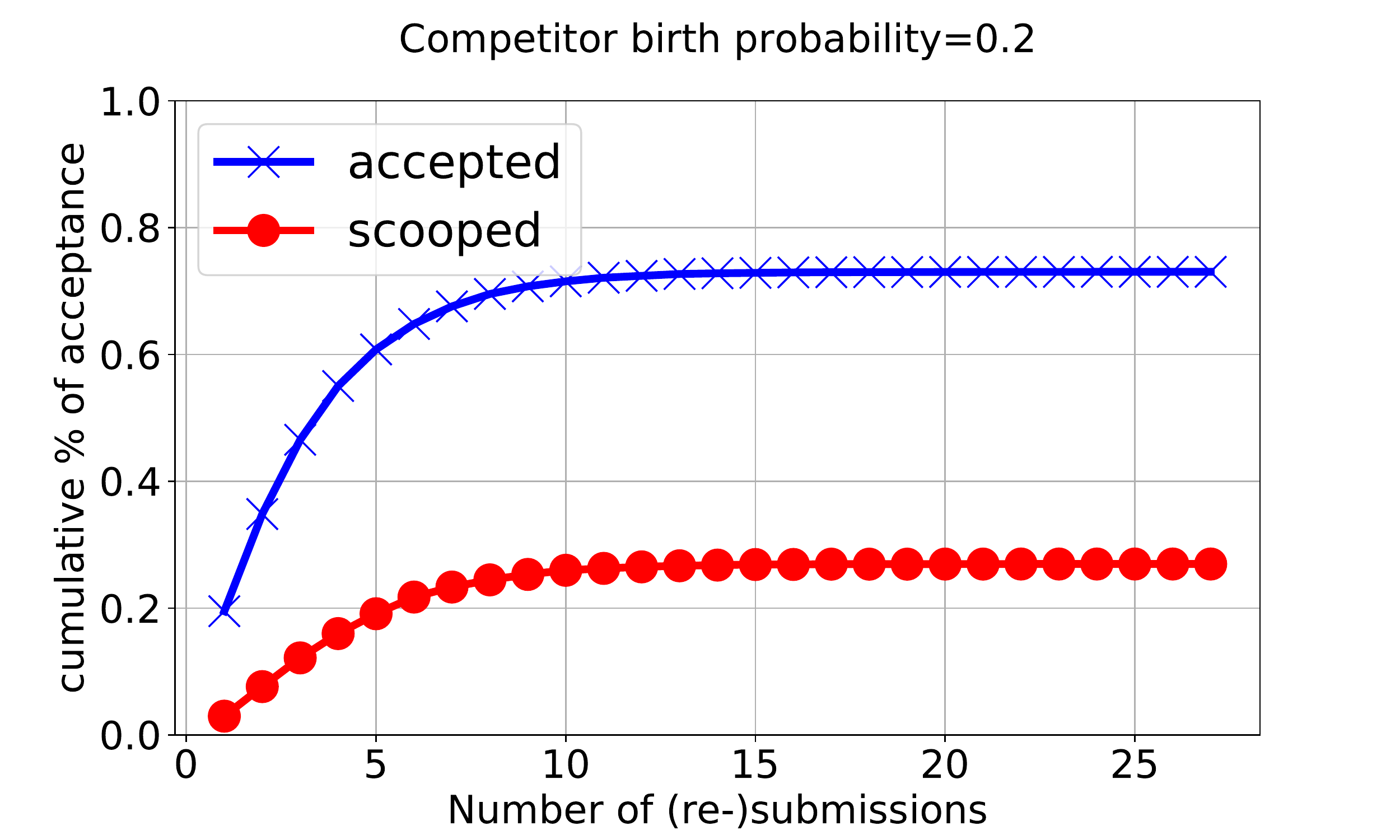}
 \includegraphics[width=0.32\linewidth]{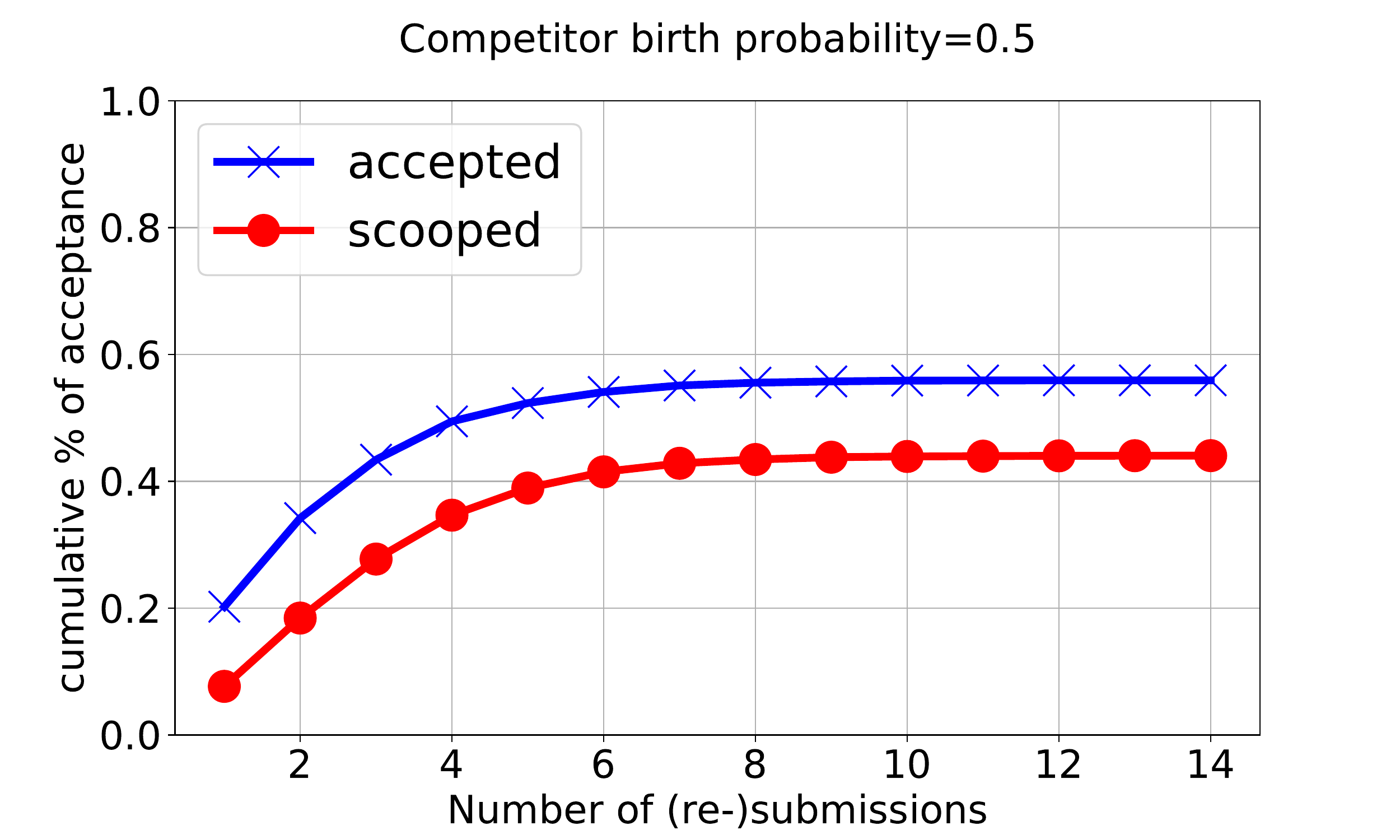}
 \includegraphics[width=0.32\linewidth]{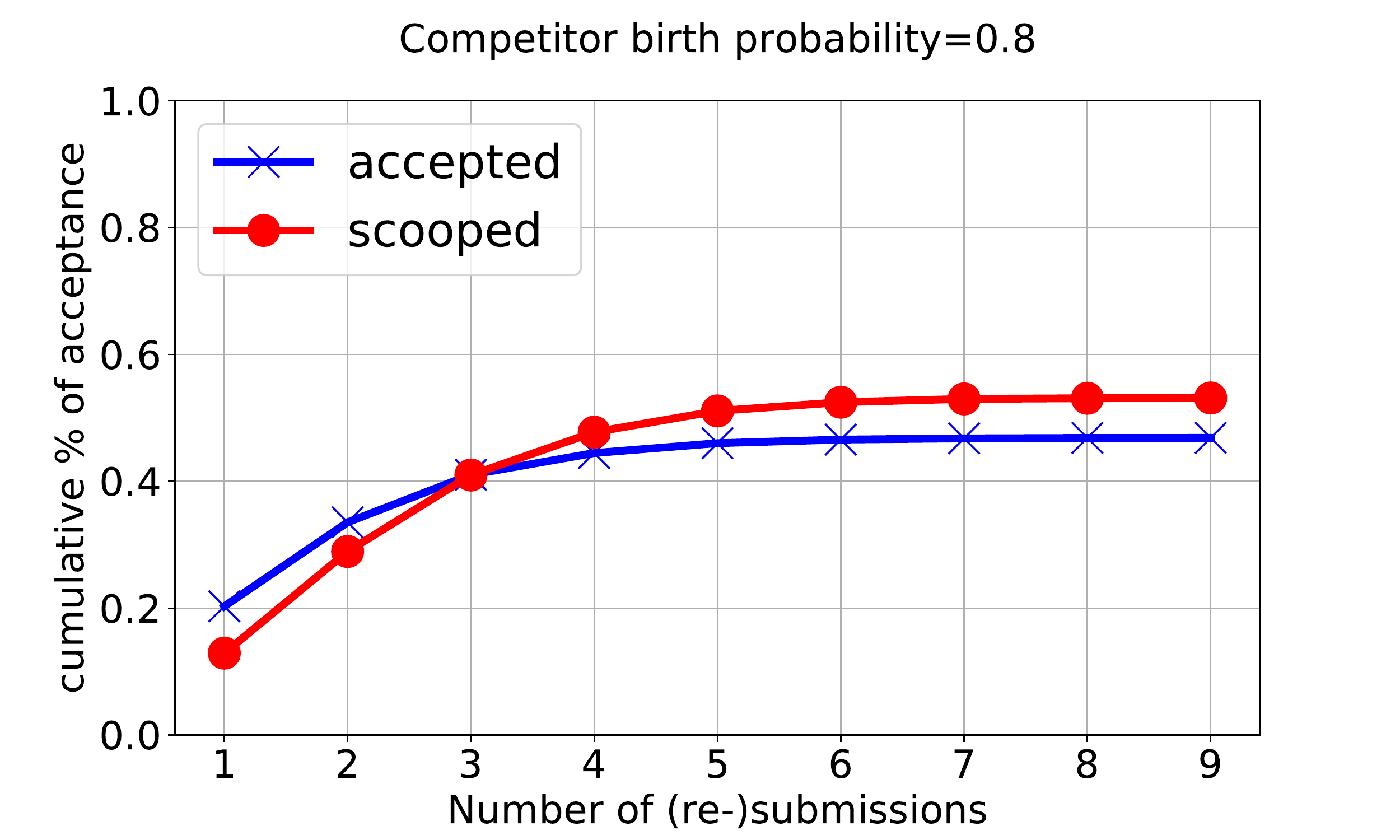}
 \caption{Monte-Carlo simulation (10k runs) of NeurIPS paper acceptance for the different competition levels, with probability 0.2 a paper is accepted. If not accepted, with probability $P_{\textit{comp}}$, then a similar idea comes to another researcher and she submits it to NeurIPS. If the competitors' paper is accepted earlier, the outcome is "scooped". The paper is resubmitted until "accept" or "scooped".} %
 \label{fig:acceptance-vs-scooping}
\end{figure}

The “wait 3-6 months” argument does not account for the research "arms race," meaning that a paper may be scooped before acceptance. By "scooping" we mean that another researcher independently comes up with the same or very similar idea and publishes it earlier.
We modeled the paper submission as a Markov process, for details see Figure~\ref{fig:acceptance-vs-scooping}.
“Wait 3-6 months” is an understatement -- it is more like "wait 1-3 years and pray to be lucky." 
In addition to independent re-discovery, a similar idea could be published by the reviewer of the submission. Not because of malicious intent -- most reviewers respect the strict confidentiality of the review process -- but because of unconscious idea generation, triggered by reading the submission. 

If an ECR waits for the paper to be accepted and it gets scooped, then there is no way to prove that they had the original idea months or years before. This is extremely unfair and painful, especially if the idea becomes influential and highly cited.

On the other hand, if the researcher uploads the work on arXiv, they get a priority timestamp and can start to get cited. Even if someone else presents a similar idea at a conference first, the affected ECR is in a much better position with arXiv.

One of the examples of protection by arXiving is a story about SiLU~\cite{Silu2016} vs Swish~\cite{swish2017} activation, which are exactly the same function. SiLU was first proposed in an arXiv paper in June 2016 by Hendrycks and Gimpel and rejected from ICLR 2017~\cite{silu-reject}.
Later, in 2017, the same activation was proposed as Swish by a Google Brain group and it was accepted to the ICLR 2018 Workshop track~\cite{swish-accept}. The SiLU was unnoticed, while Swish gained popularity. The arXiv publication helped to establish priority and give the activation function the original name and reference, SiLU, in deep learning frameworks~\cite{silu-tf, silu-pytorch}.

\mypara{Protection against gate-keeping.} 
ECRs, especially from non-mainstream labs, may have difficulty writing papers such that they conform to the norms of the community, in terms of framing their ideas and expressing them with established vocabulary. This is even more true for people with non-ML/CV backgrounds and novel ideas. Their papers could be described as “The Puppy with 6 toes"~\cite{how-to-write-good-cvpr} and are easily rejectable. Without arXiv, review gatekeeping is a single point of failure, which is hard to pass even for experienced researchers. 

Famous and less famous examples are:

\begin{itemize}[noitemsep, leftmargin=*]
  \item{The most cited computer vision paper for decades, SIFT~\cite{SIFT1999}, took three years to get accepted, first submitted in 1997, accepted as a poster in 1999, journal version published in 2004~\cite{Krig2016}.} %
  \item{SqueezeNet~\cite{squeezenet2016} was rejected from ICLR because, the “novelty of the submission is very limited," ~\cite{squeezenet-review}, while already having wide adoption and >100 citations. Now it has almost 3000 citations on Google Scholar despite never having been officially published in a conference or journal.}%
  \item{Now famous one-cycle learning rate policy~\cite{OneCycle2018} and super convergence~\cite{smith2019superconvergence} papers by Leslie Smith. Smith released the original work on arXiv in 2015~\cite{smith2015cyclicalarxiv}.  The first paper was eventually published in WACV 2017~\cite{smith2017cyclicalwacv}.  Its follow-up work, the super-convergence~\cite{smith2017superconvergence} paper was rejected from ICLR-2018~\cite{superconv-review}. Then the fast.ai team read and implemented it in their framework. Later, in 2019, it was accepted at a defence-related conference~\cite{smith2019superconvergence} little-known in the computer vision community.}
\end{itemize}

\mypara{Distribution; arXiv has become the main "new results feed," from which people discover new work}.  A whole ecosystem has been evolving around arXiv. Services like arxiv-sanity~\cite{arxiv-sanity}, arxiv-vanity~\cite{arxiv-vanity}, Papers-with-Code~\cite{paperswithcode},  arxiv-daily-type twitter accounts and recent Papers-with-Code-arXiv integration~\cite{arxiv-blog, paperswithcode-medium} allow researchers cope with the flood of published papers and supplementary materials. Moreover, arXiv is monitored by many practitioners, who are not concerned about whether the work has been endorsed by others, if there is evidence that a method is working well in practice.

Work which is not on arXiv reaches much fewer people. There is also a difference if the paper is arXived before or after acceptance. According to Feldman~\etal~\cite{feldman2018citation}, "papers submitted to arXiv before acceptance have, on average, 65\% more citations in the following year compared to papers submitted after".

\section{Research and broader community benefit from arXiv}
\label{sec:field}

\mypara{Faster and broader distribution and better explanation.} The most obvious benefit of arXiving is the distribution of authors' research to the community. This research and ideas are often clarified and explained through interaction or blog posts. 
Such explanatory and presentation work is especially important for new ideas~\cite{hamming1997art} and is logistically much more difficult for anonymous work. 

Public code and data releases are without the administrative overhead of keeping track of anonymity at all stages. While research results can be distributed in an anonymous preprint, the need to maintain anonymity restricts feedback and code sharing.

\mypara{Open access and Funding.} While companies finance research from their own profits, most university research is done on taxpayers’ money. This brings, at the very least, an obligation to share research results in a timely manner. Relatedly, arXiv serves an important role for public access to documents for grant reports and public talks.

\mypara{Crowdsourcing review of preprints is the way of Open Science and it tackles more aspects than traditional review.} Let us recall all the discussions and critique about the GPT-2~\cite{GPT2} and GPT-3~\cite{GPT3} models by OpenAI. While the opinions on GPT2/3 and its impact itself might vary, it is hard to deny that GPT2/3 started a large community-wide discussion concerning long-term impact, and machine learning bias in general. These discussions arose on Twitter and blogs at the time that the technical report and later, code, was released for GPT-2, which was posted only as an OpenAI tech report and similarly to GPT-3, released on arXiv. Would three standard reviews from NeurIPS lead to a discussion with the same scale and impact?

\mypara{arXiv levels the playing field and reduces famous labs' advantages.}
Famous labs have experienced researchers, especially in writing papers, huge hardware resources, etc. Thus, someone from a hypothetical FAANG\footnote{This acronym refers to the most prominent tech companies: Facebook, Amazon, Apple, Netflix, Google.} lab has extensive experience in how to write a “may be boring, but hard to kill” paper~\cite{how-to-write-good-cvpr}. 

While arXiv and social media publicity could make a difference for famous labs' paper acceptance, famous labs have lots of other weapons in their arsenal such as experienced writers and other resources  affecting paper acceptance.

Removing arXiv for famous labs would not make a huge difference.
Unlike them, ECRs have comparatively less \dmytro{tools in their arsenal} and removing the option to preprint before acceptance means a much reduced ability to participate in the research community. There is a Russian saying for this situation,
"\foreignlanguage{russian}{Пока толстый сохнет худой сдохнет}”, --  "While the fat one dries, the thin one dies".%

Moreover, \textbf{ECRs are more dependent on having formal publications in their C.V.} than senior researchers. If pre-submission arXiving was not allowed at major conference, ERCs would find it very difficult, unlike senior researcher, to trade early publication, code release and impact for a missed opportunity to publish at a top conference.

\section{The single blind review is not much worse than double blind.}
\label{sec:fairness-single}

\mypara{Single-vs-double blind studies -- the bias reduction is small.}
The research quoted in some anti-arXiv arguments are studies by Tomkins~\etal~\cite{Tomkins12708} about WSDM-2017 and the study by Bharadhwaj~\etal~\cite{bharadhwaj2020deanonymization} by about ICLR 2019-2020. 
Before delving into details, let us summarize both. They report, via sophisticated statistics, that deanonymization could change the chances of acceptance by 3-4 or less percentage points. In comparison, Lawrence and Cortes~\cite{nips-experiment} have found that the randomness of the review process itself is around 50\%, which is an order of magnitude bigger. We argue, that given such a level of noise, it does not make any sense to trade off all the benefits arXiv provides (Sections~\ref{sec:researcher}, \ref{sec:field}) for removing possible bias, which is 10x less that the randomness of the peer review itself. 

Besides, the first study contains methodological flaws, precisely explained in \cite{CritisismPNAS}. Specifically, Stelmakh~\etal\cite{CritisismPNAS} have shown that "the test used by Tomkins~\etal~\cite{Tomkins12708} can, with probability as large as 0.5 or higher, declare the presence of a bias when the bias is in fact absent (even when the test is tuned to have a false alarm error rate below 0.05)." Moreover, "two factors – (d) asymmetrical
bidding procedure and (e) non-random assignment of papers to referees – as is common in peer-review procedures today may introduce spurious correlations in the data, breaking some key independence assumptions and thereby violating the requisite guarantees on testing." We point the reader to~\cite{CritisismPNAS} for more details and focus on the more recent study by Bharadhwaj~\etal~\cite{bharadhwaj2020deanonymization}.

The results can be simplified as follows. Bharadhwaj~\etal have found a correlation (authors specifically emphasize that the study does not claim any causality) between pre-acceptance arXiv release  and acceptance rate. Specifically, releasing a paper on arXiv might increase its chances of acceptance up to 5 percentage points in the best case for almost everyone, or decrease your chances by 3 percentage points if the authors are totally unknown. Given that the acceptance rate at ICLR is 20-25\%, we argue that the benefits from having preprints, as stated in Section~\ref{sec:researcher}, protecting you in the very likely case of rejection, is much more important for an ECR than a possible 3 percentage point rejection risk increase.

\mypara{Authors should not be the only ones, who ensure the unbiased review.}
Instead of putting all the burden (e.g. by requiring anonymous arXiving) on the authors, effort might be directed to reviewer training. The computer vision community already started to do this by hosting CVPR Tutorial on writing good reviews~\cite{good-review-tutorial2020}. 
Another direction is to adjust the review process to minimize possible biases. 
For example, the reviewers might be asked to read the paper and write the initial review without googling the paper to avoid possible influence of knowing the author names. Such process is not uncommon in grant proposal reviewing.
After the initial review, one could perform google and literature search for the related work etc.
\section{Conclusions}
\label{sec:conclusion}
Double-blind review combined with banning non-anonymous preprints does not come for free. While possibly removing  a small bias towards "big names", it introduces other biases and limitations. Such limitations are especially harmful for those who are not yet fully established.

Is it worthwhile to do arXiving early? We argue that the default research workflow should be the following: once researchers write-up the research piece, they decide if want to submit to a conference, release on arXiv or do both. 
The choice depends on the particular situation, and only the researchers themselves should decide how to share their work.

Finally -- we have likely missed some aspects of this imporant issue and we hope that this paper will be a good starting point for a community-wide discussion. We will post updated versions on arXiv.

\section*{Broader Impact}
The submission is a position paper, discussing the impact of some aspects of submission and reviewing rules on individual researchers and the research community. The topic of the paper, and thus its content, falls under the "Broader Impact" rubric. 
\medskip

\small
\bibliographystyle{plain}
\bibliography{main.bbl}

\end{document}